\documentclass[useAMS,usenatbib]{mn2e}

\usepackage{epsfig,graphicx,amsmath,amssymb}
\usepackage{times}
\usepackage{natbib}
\usepackage{multirow}

\newcommand{\avenf}{\bar{x}_{\rm HI}}
\newcommand{\fcoll}{f_{\rm coll}({\bf x}, z, R)}
\def\myputfigure#1#2#3#4#5%
{\vskip#5pt\makebox[0pt]{\hskip#2in
\includegraphics[width=#3\textwidth]{#1}}\vskip#4pt\hfill}

\newcommand\lsim{\mathrel{\rlap{\lower4pt\hbox{\hskip1pt$\sim$}}
        \raise1pt\hbox{$<$}}}
\newcommand\gsim{\mathrel{\rlap{\lower4pt\hbox{\hskip1pt$\sim$}}
        \raise1pt\hbox{$>$}}}

\newcommand{\delT}{\delta T_b}

\title[The distribution of absorbers] {The distribution of Lyman-limit absorption systems during and after reionization} \author[D.~Crociani et al.]
{D. Crociani$^{1,2}$, A. Mesinger$^3$\thanks{Hubble Fellow; email: mesinger@astro.princeton.edu}, L. Moscardini$^{1,2}$, S. Furlanetto$^4$\\
  $^1$ Dipartimento di Astronomia, Universit\`a di Bologna, via Ranzani 1, I-40127 Bologna, Italy (daniela.crociani5@unibo.it, lauro.moscardini@unibo.it) \\
  $^2$ INFN/National Institute for Nuclear Physics, Sezione di Bologna, viale Berti Pichat 6/2, I-40127 Bologna, Italy \\
  $^3$ Department of Astrophysical Sciences, Princeton University, Princeton, NJ 08544, USA \\
  $^4$ Department of Physics and Astronomy, University of California, Los Angeles, CA 90095, USA}

\begin{document}

\date{Accepted ???. Received ???; in original...}

\pagerange{\pageref{firstpage}--\pageref{lastpage}} \pubyear{2008}

\voffset-.6in

\maketitle

\label{firstpage}

\begin{abstract}

Lyman-limit absorption systems can play many important roles during and after cosmological reionization.
 Unfortunately, due to the prohibitively large dynamic range required, it is impossible to self-consistently include these systems in cosmological simulations.  Using fast and versatile semi-numeric simulations, we systematically
explore the spatial distribution of absorption systems during and following reionization.  We self-calibrate the resulting number of absorbers to
the mean free path (mfp) of the ionizing ultraviolet background (UVB), and present results at a given mfp and neutral hydrogen fraction. We use a simple optical depth criterion to identify the locations of absorbers.  Our approach is fairly robust to uncertainties such as missing subgrid structure.
Unlike at lower redshifts where the UVB is relatively uniform, at higher redshifts the fluctuations in the UVB and the HII morphology of
reionization can drive the large-scale distribution of absorption systems.  Specifically, we find that absorbers are highly correlated with the
density field on small scales, and then become anti-correlated with the UVB on large scales.  After reionization, the large-scale power
spectrum of the absorbers traces the UVB power spectrum, which can be predicted with a simple analytic extension of the halo-model. During reionization, absorbers tend to preferentially lie inside overdensities (i.e. filaments) of the recently-ionized intergalactic medium (IGM).
Absorbers may also dominate the small-scale ($k\gsim1$ Mpc$^{-1}$) 21-cm power during and after reionization.
Conversely, they smooth the contrast on moderate scales. Once the HII regions grow to surpass the mfp, the absorbers add to the large-scale 21-cm power.
Our results should prove useful in interpreting future observations of the reionization epoch.
\end{abstract}



\section{Introduction} \label{sect:1}

Recent years have been witness to impressive advances in computer modeling of the reionization epoch \citep[for a recent review,
see][]{trac2009}.  Current, state-of-the-art simulations of reionization are capable of including the interactions of dark matter (DM), baryons,
and an evolving ionizing radiation field on $\sim$ 100 $h^{-1}$Mpc scales, while still resolving atomically-cooled halos at $z\lsim10$, which are
thought to provide the bulk of the ionizing photons in the advanced stages of reionization \citetext{\citealt{trac2008, aubert2010}; for DM-only
simulations see also \citealt{mcquinn2007a, trac2007}}.  Such simulations have verified analytic models
\citep[e.g.][]{furlanetto2004}, which conjectured that on large scales reionization proceeds in an ``inside-out'' fashion, with overdense regions
(which host the majority of sources) being ionized before underdense regions.

Thus far, most of the advances in computer modeling have focused on the {\it sources} of ionizing radiation.  Current prescriptions for assigning
ionizing luminosity to host halos and dealing with feedback mechanisms are still rather crude.  Justifiably, much research effort has gone (and
continues to go) into exploration of these uncertainties \citep[e.g.][]{mcquinn2007a, mesinger2008, croft2008}. Nevertheless, we are quite confident
in our ability to predict the distribution of DM halos that host the ionizing sources.

Unfortunately, comparatively modest progress has been made in simulating the {\it sinks} of ionizing photons during reionization (with some
notable recent exceptions; see below). These can potentially play important roles: as photon sinks, they can delay reionization
\citep{ciardi2006, alvarez2010}; they can affect the topology of reionization \citep{choudhury2009}; they can dominate the ionizing photon mean
free path (mfp) and so regulate the ionizing ultraviolet background (UVB) during and after reionization \citep{furlanetto2005,furlanetto2009};
they can be responsible for a non-negligible 21-cm signal during and after reionization \citep{choudhury2009,wyithe2009a,wyithe2009b}.

The difficulty lies in the dynamic range: small-scale structure would need to be resolved, along with the complicated feedback mechanisms
regulating it\footnote{
Although current cosmological reionization simulations can resolve the Jeans length in the mean density, ionized IGM, most
of the Universe was only exposed to an ionizing background at redshifts around $z\sim10$ \citep{larson2010}.  Before then, the collapse of
non-linear structures occurred in a neutral medium with a much smaller Jeans mass.  As the small-scale structure of the baryons depends on the
local thermal {\it history} (not the instantaneous Jeans mass; \citealp{hui1997}), simulations would need to resolve the collapse of DM and
baryons on extremely small scales (e.g. the Jeans length at $z=15$ in even the mean density IGM is an impressively-small $\sim$4 comoving pc),
including self-consistent feedback mechanisms. The situation might not be quite so bleak, however, as some fiducial astrophysical models predict an early epoch of X-ray heating (or some other form of heating), which would smooth fluctuations on $T_{\rm vir}\lsim$ few$\times10^3$ K scales, if the heating was early enough and fairly homogeneous \citep[e.g.][]{furlanetto2006a,mesinger2010}. Nevertheless, even if this dynamic range was \-a\-chieva\-ble, a copious number of realizations would be needed to fully explore the dauntingly-large astrophysical parameter space.}. 
Even at lower-redshifts ($z\lsim4$) where there is a fair amount of observational data \citep[e.g.][]{prochaska1999,peroux2003a,prochaska2009}, consensus has failed to emerge on the nature of the dominant absorbers, the so-called Lyman limit systems (LLSs), with studies associating them with low-mass dark matter halos \citep[e.g.][]{katz1996,gardner2001}, galactic or high-mass dark matter halos \citep[e.g.][]{kohler2007a} or clumpy, so-called ``cold-flows'' of IGM gas streaming onto galaxies \citep[e.g.][]{keres2005}. The situation is even more complicated at higher redshifts, where even more modest HI overabundances can contribute significantly to the ionizing photon opacity.

Instead of self-consistently simulating them, photon sinks in cosmological simulations are generally taken into account via so-called ``clumping
factors'', which are a measure of the mean recombination rate inside simulation cells \citep[though see][]{ciardi2006, mcquinn2007a}.  Clumping
factors are often assumed to be spatially (and at times even temporally) uniform, though more sophisticated techniques can use higher-resolution
simulations \citep{kohler2007b} or analytic models \citep{mcquinn2007a} when assigning clumping factors to cells.  However, adding ``by hand''
these missing small-scale modes while preserving their large-scale distribution is very uncertain.  
Furthermore, because only the clumpiness of ionized gas is relevant for the recombination rate, the clumping factor approach requires assumptions that (1) the gas is already ionized and (2) that one knows a priori the distribution of the ionized and neutral regions (see e.g. the appendix in \citealp{furlanetto2005, iliev2005a}). 

Alternatively, some simulations have included discrete, self-shielded photon sinks in the form of collapsed, unresolved structures via a
subgrid approach \citep{ciardi2006,mcquinn2007a}.  These studies use the conditional excursion set formalism \citep[e.g.][]{bond1991,lacey1993}
to populate the simulation cells with unresolved collapsed objects, concluding that they can prolong the duration of reionization
\citep{ciardi2006} and have a sub-dominant role to sources in regulating the reionization morphology \citep{mcquinn2007a}.  Again, it is unclear
if the distributions of absorbers on large scales are well preserved in such subgrid models which only depend on the local cell's density.  More fundamentally, it is unclear how to
associate absorption systems with such collapsed ``minihalos'' (with virial temperatures $<10^4$ K) or with the halos hosting the ionizing
sources \citep{wyithe2009a}.

In this paper, we take a more general approach.  We {\it do not} attempt to model the progress of reionization and the accompanying role of
absorption systems.  Nor do we add missing small scale power in an effort to estimate the recombination rate inside each cell.  Instead, {\it we
focus on predicting the large-scale distribution of absorbers.}

\citet[][MHR00]{miralda2000} and subsequently \citet{furlanetto2005} have assumed that absorption systems inside the ionized IGM at $z\sim$ 2--4
can be identified with a simple density threshold: regions of roughly the Jeans scale (computed at mean density and 10$^4$K) whose density is
above the threshold can be taken as fully neutral, while below the threshold, the IGM is fully ionized.  In their appendix,
\citet{furlanetto2005} have shown that this is similar to imposing a self-shielding cutoff, $\tau \gsim 1$,  at this critical density, given a
homogeneous UVB.  However, the ionizing background even in the cosmological HII regions becomes increasingly {\it inhomogeneous} at higher redshifts
\citep{bolton2007, mesinger2009}, due to the increasing clustering of sources and the decreasing mean free path (mfp).

Therefore in this paper, we assign self-shielded absorbers to cells with an optical depth criterion which
takes into account the inhomogeneous UVB.  We use semi-numerical\footnote{By ``semi-numerical'' we mean using more approximate physics than
numerical simulations, but capable of independently generating 3D realizations.} simulations to efficiently generate realizations of the density,
halo, ionization (i.e. HII tomography), and ionizing flux fields.  These are very fast compared to numerical simulations, allowing one to
independently vary many parameters, and yet still produce HII morphologies which are in good agreement with radiative transfer simulations 
\citep{zahn2007, mesinger2007, zahn2010,
mesinger2010}.
  We stress that, lacking the dynamic range and radiative transfer, we do not presume to model the size and photoevaporation of
absorbers, two important ingredients in answering how the absorbers affect the progress of reionization.  Instead we present results
at a given mean IGM neutral hydrogen fraction and mean absorber number density that we calibrate to the associated reduction in ionizing flux.

Our approach is very similar to the recent semi-numerical work of \citet{choudhury2009}.
Motivated by the suggestion that reionization proceeds very gradually in a ``photon-starved'' regime (as suggested by the interpretation of the
$z\gsim5$ Ly$\alpha$ forest by \citealp{bolton2007}), \citet{choudhury2009} have modeled the distribution of absorption systems and their
impact on reionization morphology.  To identify absorbers, they use an excursion-set approach comparing the local recombination
rate to the time-integrated production rate of ionizing photons.  Our approach differs from theirs in several main points (see \S~\ref{sect:2}
and \S~\ref{sect:3} for more details): (1) we estimate the reionization morphology and the instantaneous ionizing flux separately;
 (2) we use a self-shielded criterion instead of a recombination criterion to
identify the location of absorbers, the latter being more sensitive to the missing subgrid small-scale power; (3) we use a self-calibrated,
general approach to take into account (albeit crudely) missing/approximate physics.

This paper is organized as follows. In \S~\ref{sect:2} we briefly summarize our semi-numerical modeling tools.  In \S~\ref{sect:3}, we present
our prescription for identifying the locations of absorption systems.  In \S~\ref{sect:4} we present our results, including statistical
descriptions of the absorber field after and during reionization.  Finally in \S~5, we summarize our conclusions.

Unless stated otherwise, we quote all quantities in comoving units. We adopt the background cosmological parameters ($\Omega_\Lambda$, $\Omega_{\rm M}$, $\Omega_b$, $n$, $\sigma_8$, $H_0$) = (0.72, 0.28, 0.046, 0.96, 0.82, 70 km s$^{-1}$ Mpc$^{-1}$), matching the five--year results of the {\it WMAP} satellite \citep{komatsu2009}.

\section{Semi-Numerical Simulations} \label{sect:2}

\begin{figure*}
\includegraphics[scale=0.5]{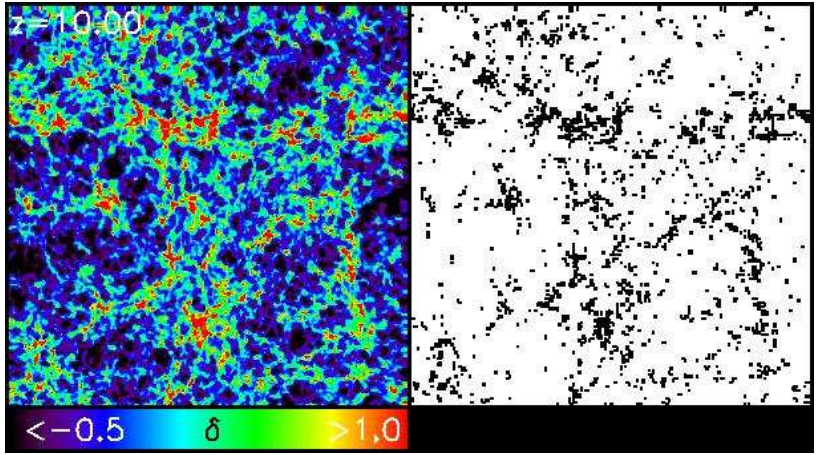}
\caption{Slices through the matter overdensity field, $\delta(\mathbf{x},z) \equiv \rho/\bar{\rho} - 1$ ({\it left panel}) and the corresponding
halo field ({\it right panel}).  Both slices are 100 Mpc on a side and 0.5 Mpc deep.
\label{fig:halo_den}}
\end{figure*}

We use the semi-numerical code DexM\footnote{http://www.astro.princeton.edu/$\sim$mesinger/Sim.html} to generate evolved density, halo, ionization and ionizing flux fields at $z=$ 10.  The halo, density and
ionization schemes are presented in \citet[][MF07]{mesinger2007}, and the ionizing flux algorithm is presented in \citet{mesinger2008}.  We refer
the interested reader to those papers for more details.  Here we briefly outline these schemes.

Our simulation box is $L=$ 100 Mpc on a side, with the final density, ionization, and flux fields having grid cell sizes of 0.5 Mpc.  Halos are
filtered out of the 1600$^3$ linear density field using excursion-set theory.  Halo locations are then mapped to Eulerian coordinates at a given
redshift using first-order perturbation theory \citep{zeldovich1970}.  This approach is similar in spirit to the ``peak-patch'' formalism of
\citet{bond1996}.  The displacement fields have an intermediate resolution of 800$^3$, in order to most efficiently use the available RAM (velocity
fields are correlated on larger scales than the density fields, thus they can afford worse resolution). The resulting halo fields have been shown
to match both the mass function and statistical clustering properties of halos in N-body simulations, well past the linear regime (MF07, Mesinger
et al., in preparation). 

The evolved (non-linear) density field is also computed in the same manner, by perturbing the 1600$^3$ Lagrangian density field with first-order 
perturbation theory.  The resulting particle locations are then binned onto a Eulerian 200$^3$ grid.  The statistical properties of the density 
fields have been shown to match those from a hydrodynamic simulation remarkably well \citep{mesinger2010}.  Specifically, the density PDF agrees 
with the numerical simulation at the percent level for over a dex around the mean density at $z\gsim7$ and roughly this cell size.  Similarly, 
the density power spectra are extremely accurate at $k\lsim5$ Mpc$^{-1}$.  We show slices through our density and halo fields in Fig. 
\ref{fig:halo_den}.

To generate the ionization field, we use excursion-set formalism \citep{furlanetto2004} which compares the number of ionizing photons produced 
in a region of a given scale to the number of neutral hydrogen atoms inside that region. Specifically, we flag ionized cells in our box as those 
which meet the criterion $\fcoll \geq \zeta^{-1}$, where $\zeta$ is some efficiency parameter and $\fcoll$ is the collapse fraction smoothed around a cell at $(\mathbf{x}, z)$ on 
decreasing scales, $R$.  The collapse fraction is computed using the resolved halo field, including halos down to a minimum mass of $10^8 M_\odot$, 
corresponding to the atomic cooling threshold at $z\sim10$. 
Note that homogeneous recombinations are implicitly included in the $\zeta$ parameter.
Also note that in this approach, unlike the recent modifications presented in \citet{zahn2010}, the IGM is considered to be a two-phase 
medium, composed of either fully-ionized or fully-neutral cells.  This is an excellent approximation on such small-scales, provided that stellar 
spectra dominate reionization (see the appendix in \citealp{zahn2010}).

We generate our ionizing flux fields following the procedure described in \citet{mesinger2008}.  We sum the contributions of halos with mass 
$M_i$ at $\mathbf{x}_i$ for which the line of sight to the considered cell at $\mathbf{x}$ only goes through ionized IGM:
\begin{equation}
\label{eq:flux}
f(\mathbf{x},z) = \frac{(1+z)^2}{4 \pi} \sum_{i} \frac{M_i\times \epsilon_{\rm ion}}{|{\bf x} - {\bf x_i}|^2} ~ e^{-|{\bf x} - {\bf x_i}|/\lambda} ~. 
\end{equation}
Here, $f(\mathbf{x},z)$ is the flux of ionizing photons at $\mathbf{x}$ (in s$^{-1}$ pcm$^{-2}$), and $\epsilon_{\rm ion}(z)$ is the the rate at which ionizing photons are released into the IGM by a dark matter halo per unit mass.  We assume a fiducial value of $\epsilon_{\rm ion}(z) = 3.8 
\times 10^{58} \Omega_b/\Omega_{\rm M} / t_H(z)$ photons $M_{\odot}^{-1}$ s$^{-1}$ which provides a good fit to the observed $z\sim6$ luminosity functions 
of Ly$\alpha$ emitting galaxies (LAEs) \citep{kashikawa2006,shimasaku2006,dijkstra2007,stark2007,mcquinn2007b}, and the $z=6$ Lyman Break 
galaxies (LBGs) (e.g. \citealt{bouwens2006}). However this fiducial normalization can be subsumed in our optical depth normalization (see \S~\ref{sect:3}) 
and is not relevant when presenting results at a fixed number density of absorbers.  What we are really after is the spatial distribution of 
flux, and its impact on the distribution of absorbers.  Hence, where applicable we present most results in terms of the normalized flux, 
\begin{equation}\label{eq:fN}
f_{\rm N}(\mathbf{x},z)\equiv f(\mathbf{x},z)/\bar{f}(z)\ ,
\end{equation}
where $\bar{f}$ is the mean value of the flux field for $\bar{x}_{\rm HI}=0$.  We chose to normalize by the flux post-reionization to facilitate 
direct comparisons at the various reionization stages.

The free parameter $\lambda$ in eq.~\eqref{eq:flux} is the assumed mean free path of the UV photons in the ionized IGM.  We use the mean free 
path to self-calibrate the number of absorbers with the flux fields, as described in \S~\ref{sect:3}.  There is considerable 
disagreement among present observational estimates of the mfp at the highest redshifts accessible to such observations, $z\sim4$ 
\citetext{\citealp{storrie1994, stengler1995, peroux2003b}, \citealp{prochaska2009}, c.f. their Fig. 14, \citealt{songaila2010}}.
Furthermore, there is no reason to expect 
that the empirically-derived redshift evolution of $\lambda$ in these studies should extend to higher redshifts. Therefore, we try to be general in our 
conclusions, and show results from two fiducial values of $\lambda$ = 10 and 20 Mpc. 

For our $\lambda=$ 10 (20) Mpc cases we obtain $\bar{f}\simeq 2.5 ~ (4.3)\times 10^{4}$ ionizing photons s$^{-1}$ pcm$^{-2}$, which corresponds to the mean ionization rate $\bar{\Gamma}\simeq 2 ~ (3.3)\times 10^{-13}$ s$^{-1}$.
  Note that in addition to matching the LAE and LBG luminosity functions as mentioned above, our procedure results in values of $\bar{\Gamma}$ which are consistent with recent analysis of the $z\sim$ 5--6 Lyman-$\alpha$ forest, provided $\bar{\Gamma}$ does not evolve much from $z\sim6$ to $z\sim10$ (e.g. \citealt{bolton2007}; note that $\bar{\Gamma}$ already was found not to evolve much from $z\sim3$ to $z\sim6$). For more detailed analyses of the ionizing flux distributions, we refer readers to \citet{mesinger2008}.

Finally, we stress that our flux fields, unlike those implicitly computed in \citet{choudhury2009}, take into account the complex morphology of 
the HII regions.  HII bubbles are not very spherical in the advanced stages of reionization \citep 
[e.g.][MF07]{mcquinn2007a, trac2007}.  Therefore, many sources within the mfp will still not contribute to the local ionization rate, since they 
do not have a ``clear line of sight'' to the cell.  Instead, these photons will go into expanding the HII regions and altering their morphological structure.

\section{Identifying  self-shielded absorption systems}\label{sect:3}

Using the modeling tools summarized in the previous section, we compute the density, halo, HII morphology, and ionizing flux fields.  The next step therefore is to identify the likely locations of self-shielded absorption systems.  As mentioned in the introduction, we do this with an optical depth criterion, which takes into account the inhomogeneity of both the density field and the UVB.  Simulation cells with 
\begin{equation}\label{eq:ss}
\tau(\mathbf{x},z)\ge 1
\end{equation}
are estimated to contain enough neutral gas to host self-shielded absorption systems. Note that since we present results at a fixed number density of absorbers, the precise value of the RHS of eq. (\ref{eq:ss}) is irrelevant (see below).
The optical depth across a cell is estimated according to
\begin{equation}
\label{eq:tau}
\tau(\mathbf{x},z)= A ~ x_{\rm HI} n_H \sigma_{\rm H}{\rm d}x\ ,
\end{equation}
where ${\rm d}x$ is our cell length of 0.5/($1+z$) pMpc, and $\sigma_{\rm H}\sim 6 \times 10^{-18}$ pcm$^2$ is
the photoionization cross section of the hydrogen \citep{osterbrock2006}, $n_H(\mathbf{x},z)$ is the cell's (proper) number density of hydrogen, $x_{\rm HI}$ is the local neutral fraction computed assuming ionization equilibrium,
\begin{equation}\label{eq:2a}
n_{\rm H} x_{\rm HI} \Gamma =n_{\rm H}^2 (1-x_{\rm HI})^2~\alpha_{\rm B}\ ,
\end{equation}
where $\alpha_{\rm B}$ is the case-B recombination coefficient\footnote{The Case A value is more appropriate if the ionizing photons are absorbed inside dense, neutral systems (e.g. MHR00); however, at higher redshifts, an increasing fraction of photons get absorbed in moderate optical depth systems \citep[e.g.][]{furlanetto2006c}. In any case, the choice of recombination coefficients is not important, as it gets subsumed in the ``A'' normalization parameter.} at $T=10^4$ K \citep{osterbrock2006}, and $\Gamma(\mathbf{x},z) =f(\mathbf{x},z) \sigma_{\rm H}$ is the local ionization rate computed using eq. (\ref{eq:flux}).

The parameter, $A$, in eq. (\ref{eq:tau}) is used to roughly (and homogeneously) incorporate missing/uncertain physical components of our approach (e.g. the absorber cross-section; see below).  We use it to self-calibrate the resulting number density of absorbers according to 
\begin{equation}
\label{eq:ss_mfp}
\lambda=\Big[\sqrt{2}n_{\rm SS}(A){\rm d}x^2\Big]^{-1}\ ,
\end{equation}
where $n_{\rm SS}(A)=N_{\rm SS}(A)/L^3$ is the number density of the self-shielded regions and $N_{\rm SS}(A)$ is the number of self-shielded pixels found through the self-shielding criterion. We self-calibrate our results by choosing $A$ such that the RHS of eq. (\ref{eq:ss_mfp}) matches the mfp\footnote{Note that eq. (\ref{eq:ss_mfp}) directly normalizes the resulting {\it number density} of absorbers, not the mfp.  If absorbers are inhomogeneously distributed, the actual mfp can be different than predicted by the above equation.} used to calculate the flux fields, i.e. eq. (\ref{eq:flux}).  For better statistics, this calibration is done in a fully-ionized universe; therefore the free-parameter $A$ is the same for all values of $\avenf$ and only depends on the mean separation of absorbers\footnote{Since $A$ is related to the intrinsic self-shielding properties of our simulation cells, it should be much more sensitive to the mfp than to $\avenf$.  
Furthermore the above calibration only makes sense if the attenuation of the ionizing background is indeed dominated by the number density of absorbers; i.e. when the ionizing photons are no longer absorbed by the HII bubble walls.  This is only the case in the advanced stages of reionization, when the HII bubbles grow to be larger than the mfp \citep{mesinger2008,furlanetto2009}.  In the early stages of reionization before this transition, the ionizing background is instead dominated by the HII morphology and is therefore insensitive to the mfp in the HII regions.}. 

Our cell size roughly matches the Jeans mass in the ionized IGM, thus crudely mimicking the effects of Jeans smoothing.  However, as noted above, small-scale fluctuations are not set by the instantaneous Jeans mass, but instead depend on the local thermal history and their state of evolution when first exposed to a UVB.  Therefore it is very likely that, like all cosmological reionization simulations, we are missing subgrid structure.
 Processes such as the subgrid recombination rate and the photoevaporation rates of absorbers are indeed very sensitive to the detailed profiles
of the ionized and neutral gas \citep[e.g][]{iliev2005b,ciardi2006,mcquinn2007a,choudhury2009}.  However, in this paper
we are merely concerned with identifying the locations of self-shielded systems.  For this task, the simple criterion of ``there is enough
neutral gas in the cell'' of eq. (\ref{eq:ss}) is likely a decent indicator that there are one or more absorption systems inside the cell (c.f.
the appendix of \citealp{furlanetto2005}). Furthermore, any homogeneous dependence of our results on the missing small-scale power (such as the
absorber cross-section) is incorporated via our calibration parameter $A$.  We
obtain $A\sim 3$ (4) for our $\lambda=10$ (20) Mpc models.

Unfortunately, there is an inconsistency in the above self-calibration.  As we shall see below, the distribution of absorption systems at high
redshifts is likely very inhomogeneous.  This makes the smooth flux attenuation term, $\exp[-r/\lambda]$, only a rough approximation.  Such a
smooth attenuation is more naturally interpreted as being due to the integrated column density of low-$\tau$ systems (i.e. the Ly$\alpha$
forest).  If these more common systems did in fact dominate the absorption of ionizing photons at high redshifts, then our assumption of a global mfp is quite reasonable.
However, if the flux was dominated by such low-$\tau$ systems, then our calibration in eq. (\ref{eq:ss_mfp}) would be unnecessary, since the flux
is not sensitive to self-shielded absorbers.  Then our results would best be interpreted at a fixed number density of absorbers, with the absorber abundance unrelated to the attenuation of the UVB.  Interestingly, \citet{furlanetto2005} find that at $z\gsim6$, the contribution of self-shielded
LLSs and the Ly$\alpha$ forest to the UV optical depth is roughly comparable, though this estimate is model-dependent and only applied to the
post-reionization regime.  Indeed, since our values of $A$ are somewhat higher than unity, it is possible that some of the absorption systems we
flag are in fact lower-$\tau$ cells more appropriate for the Ly$\alpha$ forest (see eq. \ref{eq:tau}).  An improvement to our approach would be
to assign parameterized cross-sections to each absorber and iteratively compute the UVB (which could have additional smooth attenuation from low-$\tau$ systems) and absorber fields until the two converge.  In this way the UVB and absorber distributions would be self-consistent. However this is both time-consuming and of dubious benefit, given our present ignorance of absorber properties at high-redshift.  Given the above
uncertainties, one should keep in mind that our quantitative estimates below are less robust than our qualitative ones.

Although we allow absorbers to be hosted by the same cells hosting ionizing sources, we do not explicitly model the HI content of the interstellar medium (ISM).  Resolution requirements and detailed feedback
processes make it very difficult to simulate a statistical sample of such absorbers.  At lower redshifts, high opacity systems such as Damped Lyman alpha systems (DLAs) are
indeed associated with galactic mass dark matter halos \citep{wolfe2005}.  On the other hand, low opacity systems such as the Lyman $\alpha$ forest
are associated with the filamentary structure of the web \citep[e.g.][]{schaye2001}.  At $z\sim5$, the total UV opacity is dominated by the
intermediate opacity LLSs \citep{miralda2003,peroux2005}, with the nature of these systems being the least understood.  As already mentioned, the
IGM can self-shield much easier at higher redshifts \citep{furlanetto2005}.  Therefore it is likely that at the epochs we consider, the total UV
opacity is in fact dominated by absorbers residing in the IGM, which are simpler to model.

Finally, we point out that our method, unlike the one in \citet{choudhury2009}, computes the reionization morphology separately from the absorber
locations.  The overall reionization morphology depends on the complete ionization history, whereas the absorber distributions are more sensitive
to the instantaneous ionizing flux.  
Indeed, using large-scale, DM-only simulations with radiative transfer \citet{mcquinn2007a} find that absorption systems have a small impact on reionization morphology, even with extreme assumptions, such as a small mfp and no photoevaporation (see the bottom panel of Fig. 10 in \citealt{mcquinn2007a}).  Similarly, shadowing effects from the absorbers should be small since ionizing sources are comparably very abundant with many sources from many directions contributing to the local ionizing background. Nevertheless, we stress that treating absorption systems separately from the HII morphology is an {\it assumption} of our method, {\it not} a result.

\section{Results}  \label{sect:4}


\subsection{Absorber Distributions After Reionization}

Here we briefly discuss our results post-overlap, i.e. when $\bar{x}_{\rm HI}=0$\footnote{We present results in this work at fixed $\avenf$, which is computed {\it without} the contribution from the HI inside the absorption systems.  This is the most natural way to highlight the additional impact of absorption systems, and it also avoids having to assume how much of the absorber's gas is actually neutral.}.  This is the simplest case we consider, since the HII morphology has
no impact on the results.  Therefore it is a good place to start and build some physical intuition, before proceeding to the reionization epoch.

\begin{figure*}
\includegraphics[scale=0.4]{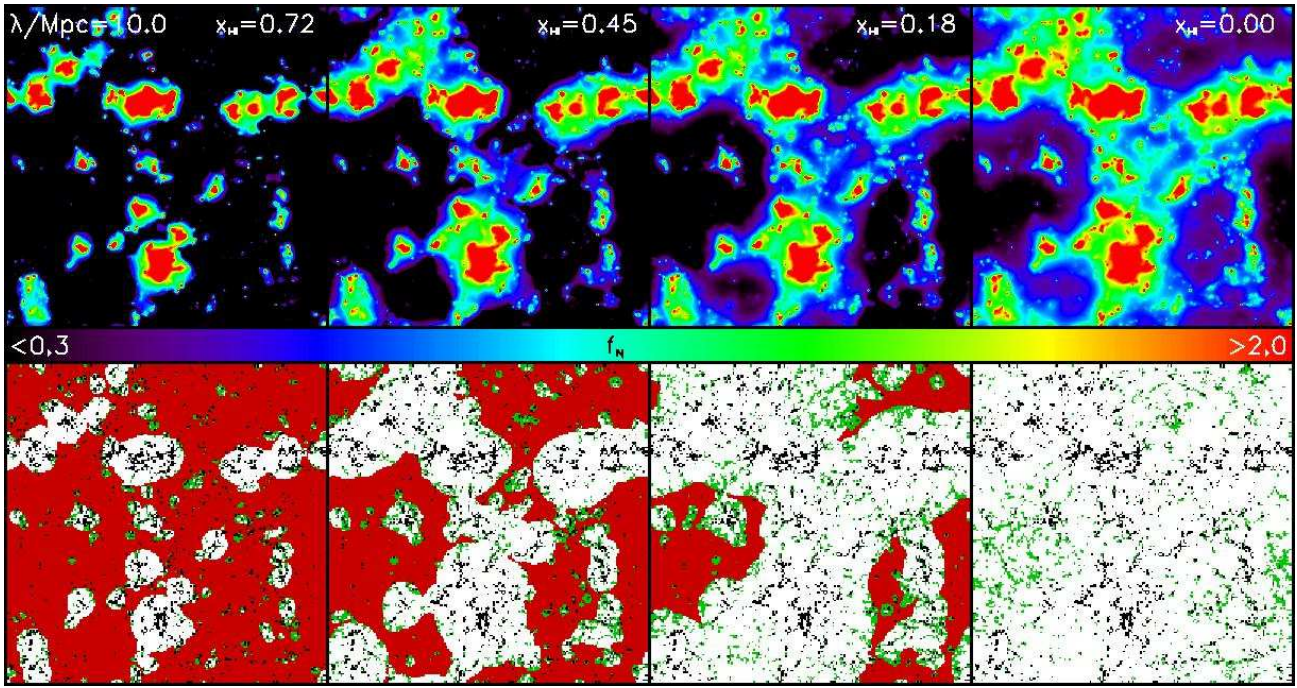}
\caption{Upper panels: the normalized flux field calculated assuming $\lambda=10$ Mpc, for $\avenf$ = 0.72, 0.45, 0.18, 0 (left to right). 
Lower panels: the corresponding ionization fields. Neutral regions are shown in red, while ionized regions are shown in white.  Halo locations are marked with black points, while absorber locations are marked with green points.
All slices are 100 Mpc on a side and 0.5 Mpc deep and correspond to the same region shown in Fig.~\ref{fig:halo_den}.\label{fig:pics10}}
\end{figure*}

\begin{figure*}
\includegraphics[scale=0.4]{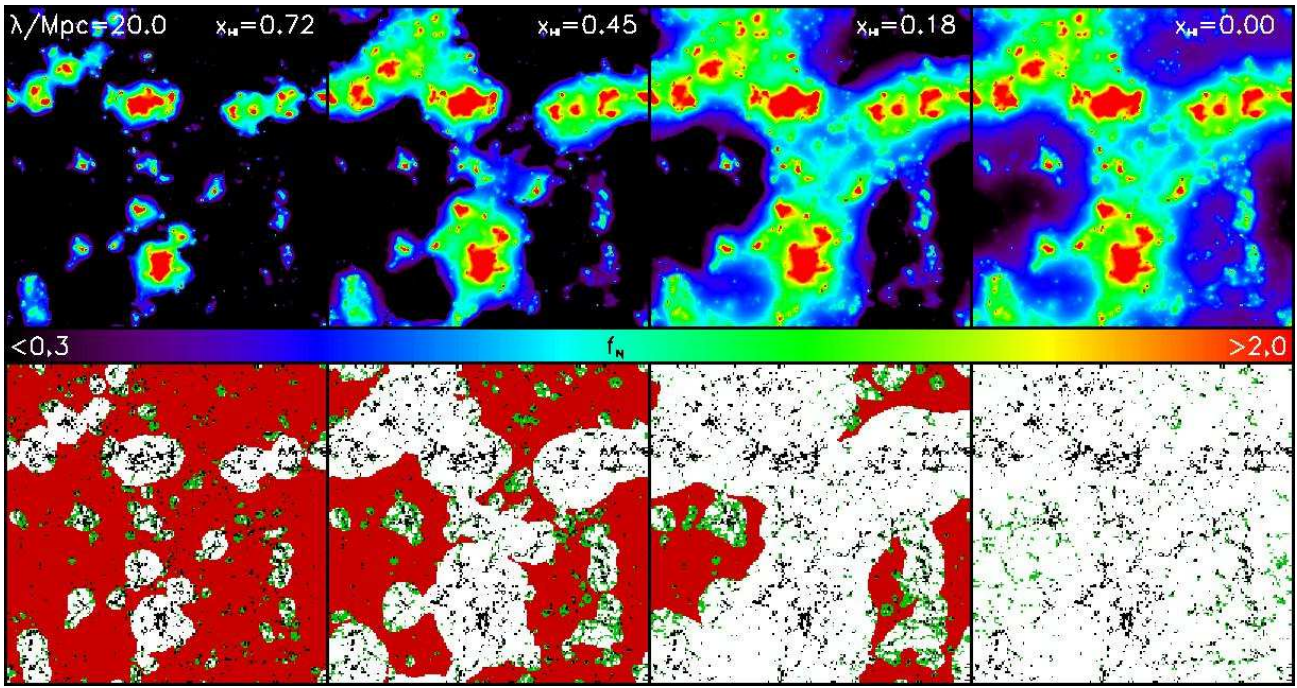}
\caption{Same as Fig.~\ref{fig:pics10}, but assuming $\lambda=20$ Mpc.
\label{fig:pics20}}
\end{figure*}

In Figures \ref{fig:pics10} and \ref{fig:pics20}, we show slices through the flux ({\it top panels}) and ionization ({\it bottom panels})
fields, generated assuming $\lambda=$ 10 Mpc (Fig.~\ref{fig:pics10}) and $\lambda=$ 20 Mpc (Fig.~\ref{fig:pics20}).
The right-most columns correspond to the post-reionization regime.

We see that even post-reionization, the UVB is spatially inhomogeneous, with the inhomogeneity increasing with decreasing mfp.  Even on large, $\sim$10
Mpc scales, spatial fluctuations in the UVB can be a factor of a few around the mean; on smaller scales, fluctuations can even be a factor of ten
around the mean \citep{mesinger2008,mesinger2009}.  This is in sharp contract to the moderate and low redshift regime ($z\lsim4$), where spatial
fluctuations contribute only a few percent of the mean value of the UVB \citep{zuo1992,fardal1993,gnedin2002,meiksin2004,croft2004,bolton2006}.

These fluctuations in the UVB indeed have a strong impact on the distribution of absorbers. Looking at the corresponding distributions of
absorbers (shown as green dots in the bottom panels of Figures \ref{fig:pics10} and \ref{fig:pics20}), one can see that they preferentially lie
in regions with a weaker UVB, far from large clusters of ionizing sources (shown as black dots).  This effect is even more pronounced when the
mfp is larger.  As a result, the absorbers themselves are spatially inhomogeneous. If this absorber inhomogeneity was self-consistently included in our UVB calculation, it would further increase its fluctuations.

On the other hand, some absorbers are seen to lie close to ionizing sources.  These absorbers likely lie in overdense regions, where the high gas
density manages to shield them from the increased ionizing radiation.  Note that the optical depth scales roughly as $\tau \propto (\delta+1)^2 f^{-1}$.

\begin{figure*}
\includegraphics[scale=0.7]{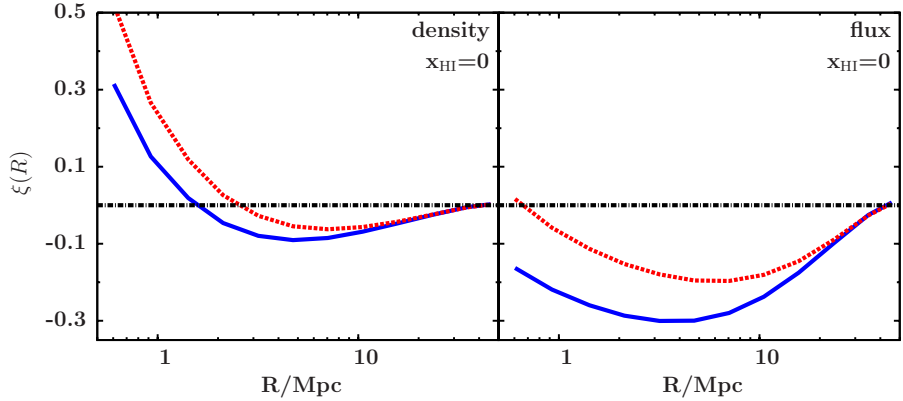}
\caption{The cross-correlation between the self-shielded regions and $\delta$ (left panel) or $f_{N}$ (right panel), at $\avenf=0$.  The blue solid and the red dashed curves correspond to $\lambda=10$ and 20 Mpc, respectively.}
\label{fig:ccc}
\end{figure*}

To quantify these trends in Fig.~\ref{fig:ccc}, we show the cross-correlation function, $\xi(R)$, of the
binary absorber field with the density ({\it left panel}) and the flux ({\it right panel}) fields.  This figure confirms the qualitative intuition from above.
At small scales, absorbers are highly correlated with overdensities.  This correlation increases as the background becomes more uniform and
there are fewer absorbers.  However, on large scales, absorbers become {\it anti-correlated} with the density field instead lying in large-scale underdense
regions where the UVB is weak.  This effect is even more prominent in the strong anti-correlation of the absorbers and the UVB on large-scales. Note that our boundary conditions which suppress modes larger than the box size bring the cross-correlation back
to zero as one approaches the box size.

Therefore scales are very important in understanding the distribution of absorbers.  On small scales, the density dependence of the optical depth
dominates, and the density field determines the absorber locations.  On the other hand, on large scales it is the ionizing flux which governs the
locations of absorbers.  This is expected since the UVB is correlated on larger scales than the density field.  The scale at which this
transition happens and the strength of the (anti-)correlation depend on the mfp.  Thus, our results argue against the common practice of adding
substructure to ionization cells based on just the local density; long-wavelength modes should also be considered.
In particular, the sources are much more biased than the absorbers, so fluctuations in the ionizing background -- which extend to large scales around a clump of sources -- dominate the variations in the absorber population.

\subsection{Absorber Distributions During Reionization}

Now that we have a better understanding of the scale-dependent absorber distributions post-reionization, we explore how these are affected by the
HII morphological structure during reionization.  Reionization morphology further modulates the ionization field, as can be seen in the top
panels of Figures \ref{fig:pics10} and \ref{fig:pics20}.  From the bottom panels of these figures, we confirm that the absorbers
preferentially cluster in regions of weak flux.  During reionization, this translates to the edges of HII regions.
When the process is viewed at a fixed $\lambda$, the absorbers essentially ''move-in'' to the newly-ionized regions.  However, inside those
large-scale regions, absorbers will preferentially be located in overdensities, i.e. filaments (see also Fig. \ref{fig:delTmaps}).

\begin{figure*}
\includegraphics[scale=0.53]{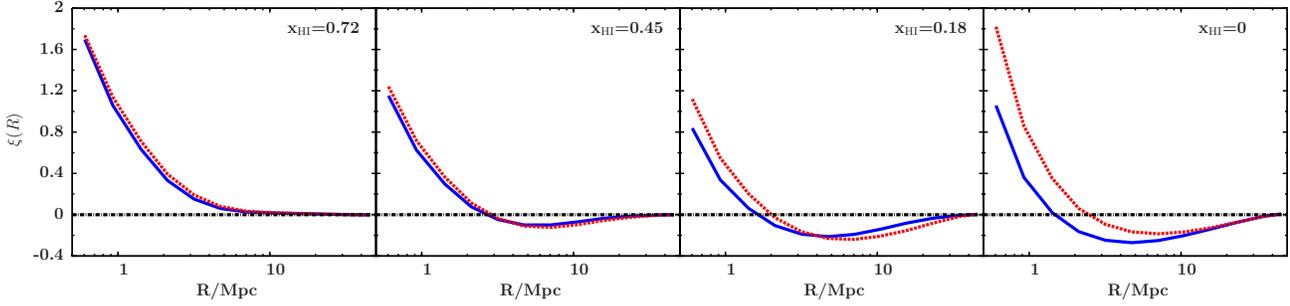}
\caption{The cross-correlation between the absorption systems and the halo field at $z=10$, for all the neutral fractions considered in this work. The blue solid and the red dashed curves refer to $\lambda=10$ and 20 Mpc, respectively.
\label{fig:halo_corr}}
\end{figure*}

These trends can also be seen more quantitatively in Fig. \ref{fig:halo_corr}, where we plot the cross-correlation of the halo and absorber
fields for $\avenf=$ 0.72, 0.45, 0.18 and 0, left to right.  The halo field is computed on the same resolution, 200$^3$ grid, with cells
containing halos flagged as `1', and those not containing halos flagged as `0'. We see that
early in reionization, when the HII regions are smaller than the mfp, there is only a positive correlation between the halos and absorbers.  This
is understandable since by definition absorbers must reside inside HII regions, and on these relatively small scales, they both correlate to the density field.  However, as the HII regions grow, the
large-scale fluctuations in the UVB become important in determining the absorber locations.  Absorbers show an increasing preference for the
newly ionized regions distant from the ionizing sources where the UVB is weak.  This manifests itself through the increasingly negative
large-scale cross-correlation as reionization progresses.

\begin{figure*}
\includegraphics[scale=0.45]{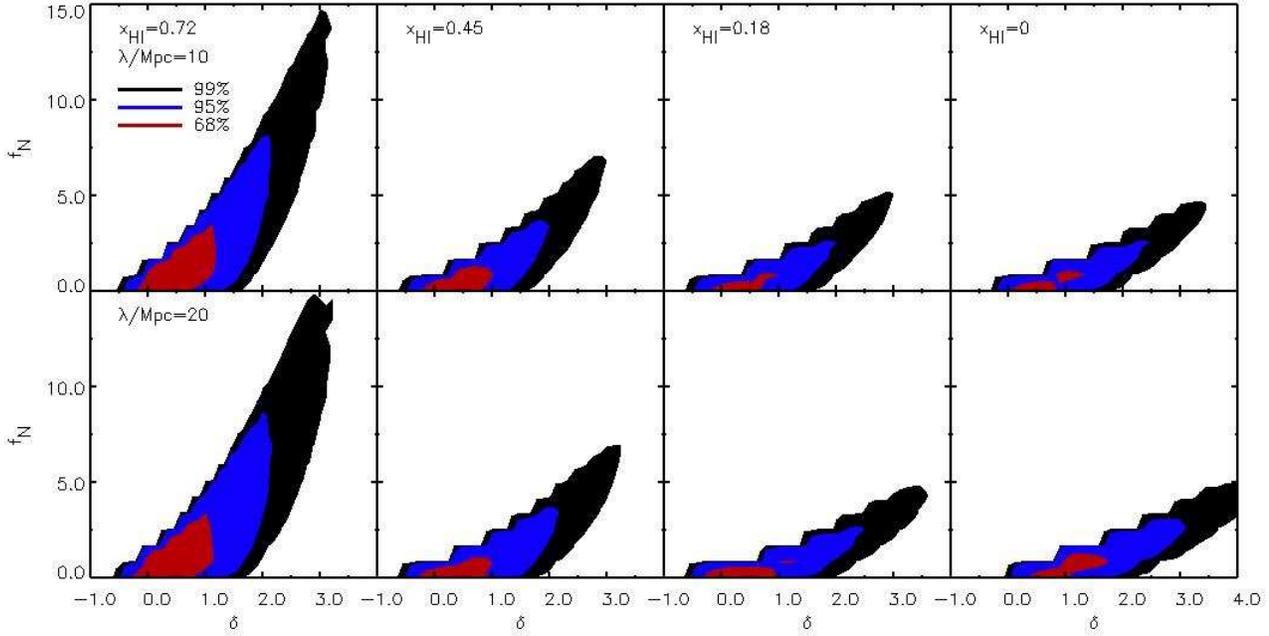}
\caption{The distribution of the absorption systems in the ($\delta$,$f_{N}$) plane.  The black, blue and red surfaces enclose 99\%, 95\% and 68\% of the absorber population, respectively. 
The upper and the lower panels correspond to $\lambda=10$ and 20 Mpc, respectively.
\label{fig:contours}}
\end{figure*}

In Fig.~\ref{fig:contours}, we plot the distribution of the absorption systems in the ($\delta$,$f_{N}$) plane.  The black, blue and red surfaces
enclose 99\%, 95\% and 68\% of the absorber population, respectively. 
The upper (lower) panels correspond to $\lambda=10$ (20) Mpc.  This figure clearly shows the degeneracy between the density and ionizing flux, as
the contours stretch from the lower left to the upper right, following $\tau \propto (\delta+1)^2/f$ isocontours.  The isocontours are flatter
for higher values of the mfp due to our choice of normalization of the ionizing flux, which means that the same $f_N$ corresponds to a higher
unnormalized value of the UVB for $\lambda=20$ Mpc than for $\lambda=10$ Mpc\footnote{Note that our normalized flux, $f_{N}$ from eq.~
\eqref{eq:fN}, is normalized to the mean value post reionization, {\it for each mfp}.  Therefore the same $f_{N}$ values correspond to the same
unnormalized ionizing flux for each $\avenf$, but are different for different values of $\lambda$.  The mean value of the UVB is of course higher
for higher values of the mfp.}.

Fig.~\ref{fig:contours} is further confirmation of our narrative above.
Early in reionization, the highly biased nature of the ionizing sources implies that ionized regions generally trace large-scale overdensities,
i.e. reionization is inside-out on large scales \citep{trac2007,mcquinn2007a,zahn2010}.  Within these large-scale regions, it is the density
field\footnote{
Note that the absorber densities in the figure are much more modest (of order unity) than in the post-reionization models of
MHR00.  MHR00 studied self-shielded systems at lower redshifts, $z =$ 2--4.   At redshift $z=10$, the mean density of the Universe is an order of
magnitude greater, and the IGM can self-shield at much smaller overdensities since $\tau\propto\Delta^2$.  Also, we study much smaller mean free
paths, which effectively means that the UVBs that we consider are much weaker, and absorbers are more common.  Finally, our semi-numerical
simulations are at a lower resolution, so we have less small-scale power than MHR00. We remind the reader that our procedure flags the locations of the absorbers but does not necessarily require that the absorbers themselves fill the entire cell.} which determines whether a cell will host absorption
systems; hence the contours in the left panels of Fig.~\ref{fig:contours} are fairly steep showing insensitivity to the UVB.  As the HII regions
grow to include large-scale voids, it is the UVB which increasingly determines whether a cell will host absorption systems; hence the contours
flatten as reionization progresses.  As a consequence, the absorbers ``move-into'' the filaments of the newly-ionized large-scale underdense
regions. This can be seen as the shift in the contours towards higher density cells. Note that cells occupying the lower right quadrant of the
figure still host absorbers, but that region of parameter space is too rare to be enclosed by the 99\% contour.

Although a direct comparison with \citet{choudhury2009} is difficult since they do not explicitly distinguish between the neutral IGM {\it
outside} HII regions and the absorbers {\it inside} HII regions, they find a similar qualitative trend of the neutral IGM remaining in
increasingly overdense regions as reionization progresses.  However, we do not find evidence for their dual peak in the density PDF of the neutral IGM at $\delta > 0$.  Although the less-dense neutral regions found by \citet{choudhury2009} could represent the neutral IGM outside HII regions, they still obtain significant neutral cells with $\delta > 0$, which is not consistent with the inside-out nature of reionization on large-scales.  The parameter studies by 
\citet{mcquinn2007a} suggest that absorbers have a small impact on reionization morphology, and that the inside-out nature of reionization on large-scales is a robust result. As already explained, this was one of the reasons we separately compute the reionization morphology and absorber locations, unlike the analytic procedure of \citet{choudhury2009} which combines the two calculations.
We stress that when the UVB is inhomogeneous, the distribution of absorbers is best represented in the ($\delta$,$f_{N}$) plane, and not marginalized over just the density.

\begin{table*}
\begin{tabular}{cccc}
\hline
$x_{\rm HI}$ & $\lambda$/Mpc & $x_{\rm aN}$ & $x_{\rm a}$\\
\hline
\multirow{2}{*}{0.72} & 10 & 0.22 & 0.06\\
		      & 20 & 0.27 & 0.08\\
\hline
\multirow{2}{*}{0.45} & 10 & 0.14 & 0.08\\
		      & 20 & 0.14 & 0.08\\
\hline
\multirow{2}{*}{0.18} & 10 & 0.09 & 0.08\\
		      & 20 & 0.05 & 0.04\\
\hline
\multirow{2}{*}{0} & 10 & 0.03 & 0.03\\
		   & 20 & 0.02 & 0.02\\
\hline
\end{tabular}
\caption{The fraction of simulation cells containing absorbers, normalized by the HII filling factor ($x_{\rm aN}$), and not normalized by the HII filling factor ($x_{\rm a}$).
\label{tab:1}}
\end{table*}

The above-mentioned trends can also be seen Table \ref{tab:1}, where we show the fraction of cells containing absorbers, normalized by the HII filling factor ($x_{\rm aN}$), and unnormalized ($x_{\rm a}$).
Indeed the abundance of absorption systems inside the ionized IGM, shown by $x_{\rm aN}$ is much higher in the early stages of reionization.  Given that reionization is inside-out on large scales, this lends further confirmation that the density field dominates absorber locations early in reionization.

\begin{figure*}
\includegraphics[scale=0.53]{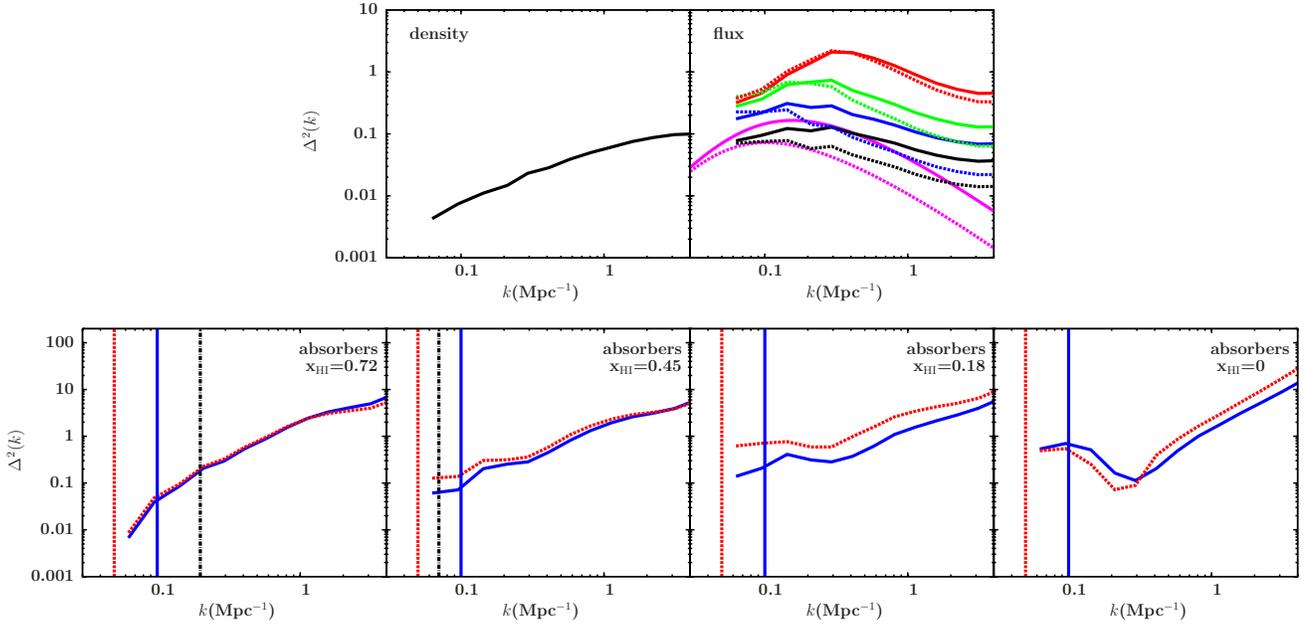}
\caption{Top panels: the power spectra of the matter fluctuations $\delta$ (left) and the UVB (right). The black, blue, green and red curves refer to $\bar{x}_{\rm HI}=0, 0.18,0.45$ and 0.72, respectively, for $\lambda=10$ (solid
lines) and 20 Mpc (dashed lines). The magenta lines in the right panel correspond to the $\bar{x}_{\rm HI}=0$ analytic model of \citet{mesinger2009} in the large-scale limit; see text for details.
Lower panels: the power spectra of the absorbers, for $\lambda=10$ (blue solid lines) and 20 (red dashed lines) Mpc. The colored vertical lines correspond to $k=1/\lambda$, while the black dot-dashed vertical line represents the mean size of the HII regions (computed according to MF07).
\label{fig:ps}}
\end{figure*}

In Fig.~\ref{fig:ps}, we show the power spectra of our various fields.  The {\it dimensionless} power spectrum of field $i$ is taken to be $\Delta^2_i(k) = k^3/(2\pi^2 V) ~ \langle|\delta_i({\bf k})|^2\rangle_k$, where $\delta_{i}({\bf x})$ is the dimensionless fluctuation of field
$i$. The top panels correspond to the density and the UVB. The bottom panels correspond to the absorber power spectra at $\avenf=$ 0.72, 0.45, 0.18 and 0, left to right, with $\lambda=10$ (20) Mpc shown with the blue solid (red dashed) curve.

Since absorbers strongly correlate with the density field on small scales, we would expect their small-scale power spectrum to roughly follow the
density power spectrum, with some bias.  On large scales however, the absorbers anti-correlate with the flux (and density)
fields.  Therefore, their large-scale power should follow the UVB power spectrum, {\it once HII bubbles have grown large enough to allow the
absorbers to populate these modes}.  We see that these trends are indeed present in the power spectra in Fig.~\ref{fig:ps}.  There is a dip in the absorber power spectrum between these two limits where the absorbers transition from being correlated to the density field on small scales to being anti-correlated on larger scales. 

Differences in the absorber power spectra of our two mfp models only start emerging when the HII bubbles grow large enough to
surpass the mfp (the mean HII bubble scale\footnote{The mean bubble size is calculated by randomly choosing an ionized cell and recording the
distance from that cell to the edge of the HII region, along a randomly chosen direction.  These distances are then averaged together to obtain
the mean bubble size.} is shown with the black vertical line in Fig.~\ref{fig:ps}).  The $\lambda=20$ Mpc curves are in general higher during
reionization, since there are fewer absorbers, and so the fractional offsets shown as the power spectra in Fig.~\ref{fig:ps} are higher.  When
shown in dimensional units (scaled by the squared mean signal), a smaller mfp does have more power, as we shall see in \S~\ref{sec:21cm}.

Unfortunately, our simulation box is only 100 Mpc on a side, which is only a few times the mfp.  Hence, it is difficult to fully isolate the
impact of $\lambda$ from the absorber power spectra.  To gain further insight, we turn to an analytic model for the UVB power spectra. \citet{mesinger2009}
have shown that the UVB power spectra can be predicted using an analogy of the ``halo model'', i.e. as the sum of two terms, one arising from
correlations within a single source's $1/r^2$ envelope, and one arising from correlations between the locations of two sources
\citep{scherrer1991, cooray2002}.  The large-scale power corresponds to the two-source term, which traces the linear matter power spectrum on
large scales times the bias, before being truncated at $k \gsim 1/\lambda$.  We plot this two source term as the magenta curves in the flux panel of Fig.~
\ref{fig:ps}.  The analytic model confirms that the $\lambda=$ 10 and 20 Mpc curves indeed do converge on large scales.  The peak and drop at $k \gsim
1/\lambda$ (demarcated by the blue and red vertical lines) is also seen in the absorber power spectra at $\avenf=0$.  The detection of this
feature would be a novel measurement of the ionizing photon mean free path, {\it valid even if the mfp was not dominated by the self-shielded absorbers themselves}.  We defer a detailed analytic treatment of the absorber power spectra, including cross- and higher-order terms, to future
work.

\subsection{Impact of the Absorbers on the  21-cm Signal}
\label{sec:21cm}

 Some of the most important and timely observations of the epoch of reionization will come in  the form of the redshifted 21-cm line of neutral hydrogen.  Several interferometer telescopes are scheduled to become active in the near future, including the Mileura Wide Field Array (MWA; \citealt{bowman2005})\footnote{http://web.haystack.mit.edu/arrays/MWA/}, 
 the Low Frequency Array (LOFAR; \citealt{harker2010})\footnote{http://www.lofar.org}, the Giant Metrewave Radio Telescope (GMRT; \citealt{pen2008}), the Precision Array to Probe the Epoch of Reionization (PAPER; \citealt{parsons2009}), and eventually the Square Kilometer Array (SKA)\footnote{http://www.skatelescope.org/}.  The neutral hydrogen of the absorbers could have a significant impact on the interpretation of the reionization signal, as suggested by \citet{choudhury2009}.  Here we investigate this signal in our models.  Note again that we do not explicitly focus on the HI content of the ionizing galaxies with $T_{\rm vir}\gsim10^4$ K, as in \citet{wyithe2009a}.

The offset of the 21-cm brightness temperature from the CMB temperature along a line of sight (LOS) at observed frequency $\nu$ can be written
as (c.f. \citealt{furlanetto2006b}):

\begin{equation}
\label{eq:delT}
\frac{\delta T_{b}(\mathbf{x},z)}{\rm mK} \approx 9(1+z)^{1/2}x_{\rm HI}(\mathbf{x},z)[1+\delta_{\rm nl}(\mathbf{x},z)] ~ ,
\end{equation}

\noindent where we have chosen to ignore peculiar velocity effects, shown to have a relatively minor impact during the advanced stages of
reionization (\citealp{mcquinn2006}, MF07, \citealp{mesinger2010}). Equation \eqref{eq:delT} further assumes that the spin temperature of the IGM
has been heated well past the CMB temperature, which seems likely during the bulk of reionization for fiducial astrophysical models
\citep[e.g.][]{pritchard2006,furlanetto2006b,chen2008,santos2008,baek2009,mesinger2010}.

\begin{figure*}
\includegraphics[scale=0.4]{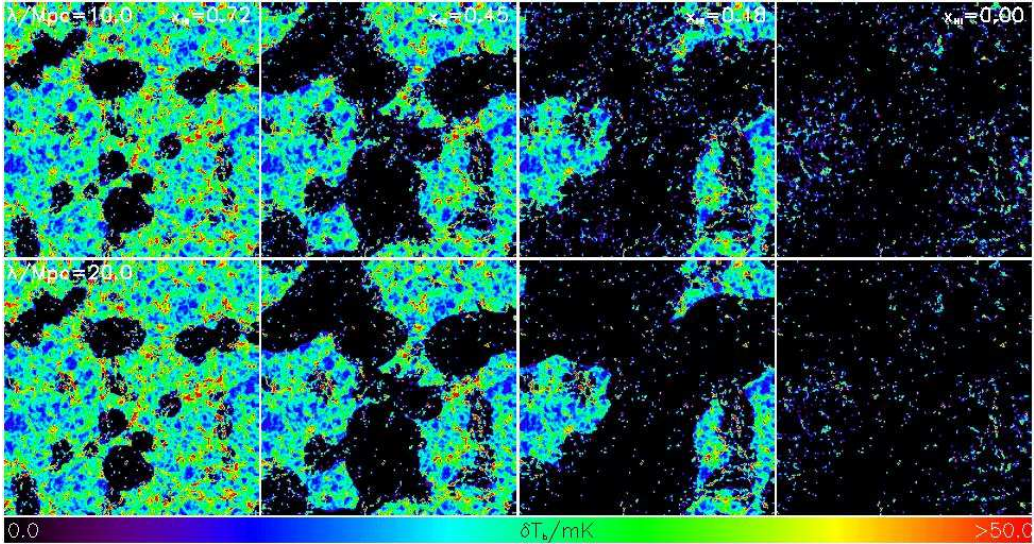}
\caption{Slices through our $\delT$ boxes, assuming $\lambda$ = 10 ({\it upper panels}) and 
20 Mpc ({\it lower panels}), and $\bar{x}_{\rm HI}=0.72,0.45,0.18$ and 0 ({\it left to right}).  All slices are 100 Mpc on a side and 0.5 Mpc
deep and correspond to the same region shown in Fig.~\ref{fig:halo_den}.
\label{fig:delTmaps}}
\end{figure*}

\begin{figure*}
\includegraphics[scale=0.5]{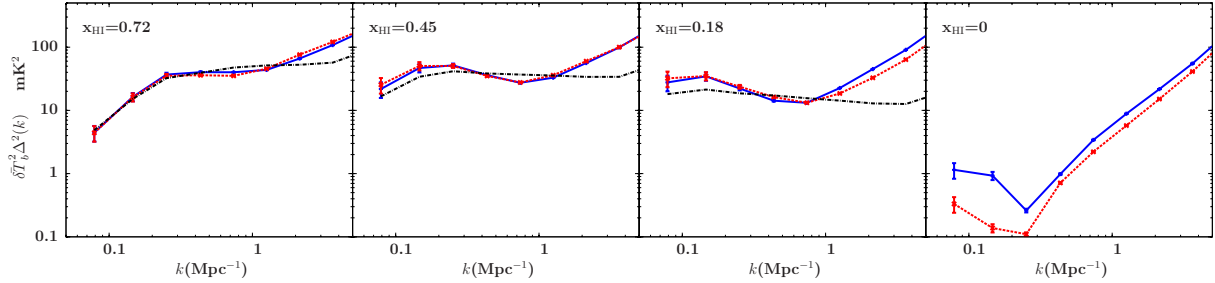}
\caption{The dimensional power spectrum of 21-cm fluctuations, estimated assuming $\lambda=10$ (blue solid lines) and 
20 Mpc (red dashed lines), for $\bar{x}_{\rm HI}=0.72, 0.45, 0.18$ and 0  ({\it left to right}).
The black dot-dashed curve corresponds to the 21-cm power neglecting absorption systems.
\label{fig:delTps}}
\end{figure*}

Slices through our 21-cm brightness temperature boxes are shown in Fig.~\ref{fig:delTmaps}, for $\avenf=0.72,0.45,0.18$ and 0 (left to right) and
$\lambda$ = 10 ({\it upper panels}) and 20 Mpc ({\it lower panels}).  Black regions correspond to the HII bubbles, with the absorbers seen to
preferentially populate the outskirts of these regions.  The filamentary structure of the neutral IGM can be seen outside the HII regions, with
the strongest (reddest) signal coming from overdensities.  Looking closely at Fig.~\ref{fig:delTmaps} one can note that indeed as the HII regions
expand into the neutral IGM, the absorbers preferentially correspond to the overdensities in the newly-ionized medium.

The corresponding {\it dimensional} 3D power spectra, $\bar{\delT}^2 \Delta^2_{21}(k)$, are shown in Fig.~\ref{fig:delTps}, with $\lambda=10$ (20)
Mpc corresponding to the blue solid (red dashed) curves.  The  black dot-dashed curve corresponds to the 21-cm power neglecting absorption
systems\footnote{As mentioned previously, $\avenf$ is computed {\it without} the contribution from the HI inside the absorption systems.  The ``total'' neutral fraction of the Universe would be $\bar{x}_{\rm HI}^{\rm tot} = \avenf + x_{\rm HI}^{\rm abs}$, where $x_{\rm HI}^{\rm abs}$ is the fraction of the IGM gas contained as HI inside absorption systems.  In our fiducial case where all of the absorber's gas is assumed to be neutral (an assumption only relevant for the quantitative predictions in this section), $x_{\rm HI}^{\rm abs} \approx x_{\rm a}$, which can be read from Table 1.}.

Neglecting absorbers, this signal progresses along the familiar story lines.
Before reionization, the 21-cm power spectrum is dominated by the matter power spectrum.  Reionization carves out dark regions in the 21-cm
signal, which initially ($\avenf\gsim0.9$) steepen the power spectrum as the first regions to be ionized are the small-scale overdensities,
equilibrating power on large scales \citep{mcquinn2007a}.  As the HII regions grow, the power spectrum is governed by HII morphology, with a
``shoulder'' feature corresponding to the bubble scale progressing from small to large scales, and stalling at the mfp scale \citep{mesinger2010}.

Including absorbers complicates this story.  Most notably, they drive small-scale power.  Therefore the 21-cm power spectrum increases at small
scales ($k \gsim 1$ Mpc$^{-1}$) where the HII bubbles would have otherwise damped the fluctuations.  On moderate scales (0.5 Mpc$^{-1}\lsim k
\lsim$ 1 Mpc$^{-1}$), the absorbers actually smooth the fluctuations somewhat by decreasing the contrast between the ionized and neutral regions.
Once the HII regions grow to be larger than the mfp, the absorber distribution (sourced by the UVB) adds to the large-scale power.

Again we note that the power spectrum is not sensitive to the mfp until the later stages of reionization.  The modulation of the UVB by a smaller
mfp and the increased number density of absorbers result in a stronger signal of the $\lambda=$ 10 Mpc model than the 20 Mpc model.

As noted by \citet{choudhury2009}, the IGM absorbers can contribute to a sizable 21-cm signal
 (though see \citealt{lidz2008}) during and post-reionization (more so than galactic HI;
e.g. \citealt{wyithe2009a}).  Overall, we find a smaller 21-cm signal than \citet{choudhury2009}, who also do not find the non-monotonic
behavior of the power spectrum noted above.  If indeed the 21-cm signal from absorbers is strong even post-reionization, this could bias the
calibration of upcoming interferometers.

We remind the reader that our procedure assigns all of the gas inside the absorber's simulation cell as being fully neutral.  This need not be
the case, as mentioned previously.  Indeed, low redshift LLSs are found to be mostly ionized \citep[e.g.][]{peroux2003b}, although absorption
systems during the reionization epoch likely have different properties.  Nevertheless, readers should note that the absorbers
could contribute less to the 21-cm signal than shown here\footnote{
In the previous sections we focus on the spatial distribution of absorbers, which should be robust to these uncertainties. 
Furthermore, the dimensionless power spectra in Fig.~\ref{fig:ps} is less sensitive to subgrid structure, since the fractional variation,
$\delta\equiv\rho/\bar{\rho}-1$ is unchanged for a constant absorber ionization fraction.}.
In fact, if we allow the absorber host cells to be partially ionized, a residual HI fraction of only 10$^{-3}$ -- 10$^{-2}$ is sufficient to achieve a lyman limit optical depth of $\tau \sim 1$.  This would decrease the absorber contribution to the 21-cm power spectrum by roughly the same amount, making it negligible. Partially ionized high-$z$ absorbers might be more common if, for example, higher energy photons can efficiently penetrate the absorbers.  
Regardless of what exactly is the neutral fraction of the absorber cells, the qualitative trends in the 21cm power spectrum discussed in this section should remain unchanged.

\section{Conclusions} \label{sect:5}

Lyman-limit absorption systems can play many important roles during and after cosmological reionization: as photon sinks, they can delay
reionization; they can regulate the UVB; they can affect the morphology of reionization; they can be
responsible for a non-negligible 21-cm signal.  Unfortunately, due to the prohibitively large dynamic range required, it is impossible to
self-consistently include these systems in cosmological simulations.

In this paper, we focus on predicting the spatial distributions of absorbers during and following reionization.  We use the efficient semi-numerical simulation code, DexM, to generate density, halo, ionization, and UVB fields.  We then use a simple optical depth criterion to identify the simulation cells expected to host Lyman-limit absorption systems.
  We self-calibrate the resulting number of absorbers to the mfp of the UVB, and present results at a given mfp and
neutral hydrogen fraction.  Our approach is fairly robust to
uncertainties such as missing subgrid structure.

We find that absorbers are strongly correlated with the density field on small scales.  However, on large-scales the absorbers instead are strongly anti-correlated with the UVB, once HII bubbles grow enough to allow absorbers to populate these modes.  Therefore we caution that using only a density criterion when assigning LLSs to cosmological simulations in fact neglects large-scale modes sourced by UVB fluctuations.

Our ionization field algorithm assumes that absorbers do not have a large impact on reionization morphology 
\citep{mcquinn2007a}.  As such, reionization in our simulations proceeds in an ``inside-out'' fashion on large scales.  Early in reionization, when the HII bubbles are small, the density field regulates the locations of absorbers.  As HII bubbles expand into the large-scale underdensities, the absorbers tend to preferentially populate the overdensities (i.e. filaments) of the recently-ionized IGM.  Therefore the absorbers effectively huddle around the edges of the HII regions.

 If the absorbers are indeed preferentially found
on the outskirts of HII bubbles, the reionization could effectively stall in the late stages.  Reionization of the remaining HI regions might
have to wait until the absorbers get photo-evaporated by the relatively faint UVB at the edges of the HII regions \citep[e.g.][]{furlanetto2009}.
Alternately, new ionizing sources could reionize the remaining HI regions from the inside.  Depending on the nature of the absorption systems
\citep{iliev2005b}, the time-scales of both processes could be fairly long: of order  $\Delta z \sim 1$.

If high-$z$ absorbers are significantly neutral, they may also dominate the small-scale ($k\gsim1$ Mpc$^{-1}$) 21-cm power during and after reionization.  They smooth the contrast on
moderate scales, and once the HII regions surpass the mfp, the absorbers add to the large-scale 21-cm power.  Since our procedure assumes all of the gas inside a simulation cell is neutral, our quantitative estimates of the absorber contribution to the 21-cm signal can be viewed as upper limits.  However, the qualitative trends should be robust.

Our results should prove useful in interpreting future observations of the reionization epoch. We also plan to extend our analysis to HeII reionization, where the sizable inhomogeneity of the UVB \citep{dixon2009,mcquinn2009} will have an even more pronounced impact on HeII Lyman-limit absorption systems.

\section*{acknowledgements}
We thank Matt McQuinn and Adam Lidz for insightful comments on a draft version of this paper. We acknowledge financial contribution from contracts ASI-INAF
I/023/05/0, ASI-INAF I/088/06/0 and ASI-INAF I/016/07/0.
 Support for this work was also provided in part by NASA through Hubble Fellowship grant HST-HF-51245.01-A to AM, awarded by the Space Telescope Science Institute, which is operated by the Association of Universities for Research in Astronomy, Inc., for NASA, under contract NAS 5-26555.
SRF was partially supported by the David and Lucile Packard Foundation and by NASA through the LUNAR program. The LUNAR consortium (http://lunar.colorado.edu), headquartered at the University of Colorado, is funded by the NASA Lunar Science Institute (via Cooperative Agreement NNA09DB30A) to investigate concepts for astrophysical observatories on the Moon.
\bibliographystyle{mn2e}
\bibliography{master2.bib}
\label{lastpage}
\end{document}